\documentclass[letterpaper, 10pt, conference]{ieeeconf}      

\IEEEoverridecommandlockouts                              

\usepackage{mathptmx} 
\usepackage{times} 
\usepackage{amsmath} 
\usepackage{amssymb}  
\usepackage{algorithm}
\usepackage{algorithmic}
\usepackage{textcomp}

\usepackage{graphicx}
\usepackage{epstopdf}
\usepackage{mathrsfs}
\usepackage{color}

\providecommand{\norm}[1]{\lVert#1\rVert}
\newcommand{\IM}{\mathrm{im}}

\newcommand{\Rank}{\mathrm{rank}}
\newcommand{\SPAN}{\mathrm{span}}
\newtheorem{Definition}{Definition}
\newtheorem{Theorem}{Theorem}
\newtheorem{Lemma}{Lemma}

\newtheorem{Corollary}{Corollary}
\newtheorem{Remark}{Remark}
\newtheorem{Notation}{Notation}
\newtheorem{Example}{Example}

\newtheorem{Problem}{Problem}

\title{
   Moment Matching Based Model Reduction for LPV State-Space Models
}

\author{Mert Ba\c{s}tu\u{g}$^{1,2}$, Mih\'{a}ly Petreczky$^{2}$, Roland T\'{o}th$^{3}$, Rafael Wisniewski$^{1}$, John Leth$^{1}$ and Denis Efimov$^{4}$
\thanks{$^{1}$Department of Electronic Systems, Automation and Control, Aalborg University, 9220 Aalborg, Denmark {\tt\small mertb@es.aau.dk, raf@es.aau.dk, jjl@es.aau.dk}}%
\thanks{$^{2}$Department of Computer Science and Automatic Control (UR Informatique et Automatique), \'{E}cole des Mines de Douai, 59508 Douai, France {\tt\small mihaly.petreczky@mines-douai.fr}}%
\thanks{$^{3}$Department of Electrical Engineering, Control Systems, Eindhoven University of Technology, P.O. Box 513, 5600 MB, Eindhoven, The Netherlands {\tt\small r.toth@tue.nl}}%
\thanks{$^{4}$Non-A, INRIA Lille - Nord Europe, Parc Scientifique de la Haute Borne, 40 avenue Halley, B\^{a}t. A Park Plaza, 59650 Villeneuve d'Ascq, France {\tt\small denis.efimov@inria.fr}}
\thanks{This work was partially supported by ESTIREZ project of Region Nord-Pas de Calais, France, the Danish Council for Strategic Research (contract no. 11-116843) within the 'Programme Sustainable Energy and Environment' under the "EDGE" (Efficient Distribution of Green Energy) research project, and the Netherlands Organization for Scientific Research (NWO, grant no: 639.021.127).}
}

\begin{document}

\maketitle
\thispagestyle{empty}
\pagestyle{empty}

\begin{abstract}

We present a novel algorithm for reducing the state dimension, i.e. order, of linear parameter varying (LPV) discrete-time state-space (SS) models with affine dependence on the scheduling variable. The input-output behavior of the reduced order model approximates that of the original model. In fact,  for input and scheduling sequences of a certain length, the input-output behaviors of the reduced and original model coincide. 
The proposed method can also be interpreted as a reachability and observability reduction (minimization) procedure for LPV-SS representations with affine dependence.

\end{abstract}

\section{INTRODUCTION}

In control applications, it is often desirable \cite{toth_book,Rugh00} to use discrete-time linear parameter-varying state-space representations with affine dependence on parameters   (abbreviated as \emph{LPV-SS  representations} in the sequel) of the form:
\begin{equation}
\label{eq:alpv_int_1}
\Sigma\left\{
\begin{split}
   x(t+1) &=& A(p(t))x(t)+B(p(t))u(t) \\
   y(t) &=& C(p(t))x(t),
\end{split}\right.
\end{equation}
where $t \in \mathbb{N}$,
$x(t) \in \mathbb{R}^{n_x}$ is the state, $y(t) \in \mathbb{R}^{n_y}$ is the output, $u(t) \in \mathbb{R}^{n_u}$ is the input, and 
$p(t)= \begin{bmatrix} p_1(t) & \cdots & p_{n_p}(t) \end{bmatrix}^\mathrm{T} \in \mathbb{P} \subseteq \mathbb{R}^{n_p}$ is the scheduling signal at time $t \in \mathbb{N}$.
Here $\mathbb{P}$ is an arbitrary but fixed subset of $\mathbb{R}^{n_p}$ with a non-empty interior, and 
$\mathbb{N}$ denotes the set of natural numbers including zero. 
The matrices $A(p(t))$, $B(p(t))$, $C(p(t))$ in \eqref{eq:alpv_int_1} are assumed to be affine and static functions of $p(t)$ of the form:
\begin{equation} \label{eq:alpv_int_1_add}
\begin{split}
A(p(t))=A_0 + \sum_{i=1}^{n_p}A_ip_i(t), &  \\
B(p(t))=B_0 + \sum_{i=1}^{n_p}B_ip_i(t),  & \\
C(p(t))=C_0 + \sum_{i=1}^{n_p}C_ip_i(t), & 
\end{split}	
\end{equation}

where $A_i \in \mathbb{R}^{n_x \times n_x}$, $B_i \in \mathbb{R}^{n_x \times n_u}$, $C_i \in \mathbb{R}^{n_y \times n_x}$ are constant matrices  for all $i \in \{ 0,1, \dots, n_p \}$.

\textbf{Contribution of the paper}
Consider a LPV-SS representation $\Sigma$ of the form \eqref{eq:alpv_int_1} and fix a positive integer $N$. In this paper, we present a procedure for computing another LPV-SS representation
\begin{equation} \label{eq:alpv_int_2}
  \bar{\Sigma}\left\{
  \begin{split}
  & \bar{x}(t+1) = \bar{A}(p(t))\bar{x}(t)+\bar{B}(p(t))u(t) \\
  & \bar{y}(t) = \bar{C}(p(t))\bar{x}(t),
  \end{split}\right.
 \end{equation}
 such that for $x(0)=0$,  $y(t)=\bar{y}(t)$ for $0 \leq t \leq N$, for all scheduling sequences $(p(0), p(1), \dots, p(N)) \in \mathbb{P}^{N+1}$ and input sequences $u=(u(0), u(1), \dots, u(N))  \in (\mathbb{R}^{n_u})^N$. 
Moreover, the state space dimension of $\bar{\Sigma}$ is smaller than or equal to the state space dimension of $\Sigma$. In other words, given an LPV-SS representation $\Sigma$ of order $n_x$ (state space dimension $n_x$) and a $N \in \mathbb{N} \backslash \{0\}$, we would like to find another LPV-SS representation $\bar{\Sigma}$ of order $r \le n_x$ which has the same input-output behavior for all scheduling and input sequences of length up to $N+1$\footnote{Note that finding a representation $\bar{\Sigma}$ with the same number of states as $\Sigma$ is in fact not necessarily useful, but it can happen that the proposed method does not allow us any other option.}. In addition, we would like the representation $\bar{\Sigma}$ to be a ``good'' approximation of $\Sigma$ in terms of input-output behavior, even for scheduling and input sequences of length greater than $N+1$ (see Remark \ref{rem:N_and_n} for what is meant by ``good'' here). 
Intuitively, it is clear that there is relationship between $N$ and $r$: larger $N$ yield a better approximation of the original input-output behavior, but they also result in larger values of $r$. 
In this paper, this relationship will be made more precise. Finally, by making use of this relation, the number $N$ can be \emph{guaranteed} to be chosen such that the resulting representation is a complete realization of the original model and it is reachable and/or observable. Therefore, the procedure stated in the present paper can also be used for reachability or observability reduction (hence, minimization) of an LPV-SS representation.

\textbf{Motivation}
  LPV-SS representations are used in a wide variety of applications, see for instance \cite{marcos_balas,verdult_verhaegen,bianchi,toth2006,dettori}. Their popularity is due to their ability to capture nonlinear dynamics, while remaining
  simple enough to allow effective control synthesis, for example, by using optimal $\mathcal{H}_2/\mathcal{H}_\infty$ control, Model Predictive Control or PID approaches.
  LPV-SS representations arising in practice, especially which arise from first-principles based modeling methods, often have a large number of states. 
  This is due to the inherent complexity of the physical process whose
  behavior the LPV-SS representations are supposed to capture.  Unfortunately, due to memory limitations and numerical issues, the existing LPV controller  synthesis tools are not always capable of 
  handling large state-space representations  \cite{WernerIFAC2014}. Moreover, even if the control synthesis is successful, large plant models lead to large controllers. In turn, large
  controllers are more difficult and costly to implement, and they often require application of reduction techniques. 
  For this reason, model reduction of LPV-SS representations is extremely relevant for improving the applicability of LPV systems.

To the best of our knowledge, the results of this paper are new. 
The tools which have been used in this paper stem from realization theory of LPV-SS representations \cite{petreczky_mercere,toth2012}. 
Similar tools were used for linear switched systems in \cite{bastugACC2014}. In fact, we use the relationship between LPV-SS representations and 
linear switched systems derived in \cite{petreczky_mercere} to adapt the tools of \cite{bastugACC2014} to LPV-SS representations. 
The method employed in this paper is related to that of \cite{toth2012}. 
The main  difference is that \cite{toth2012} requires the explicit computation of Hankel matrices of LPV-SS representations.
It should be noted that the size of the partial Hankel matrix of an LPV-SS representation increases exponentially (this will be stated more clearly in the paper, after necessary definitions are made). 
 In contrast, the algorithm proposed in this paper does not require the explicit computation of Hankel matrices, and its worst-case computational complexity is polynomial. 
 We present an example where the algorithm of \cite{toth2012} is not feasible due to the large size of the Hankel-matrix, while the algorithm of this paper works without problems. 

Regarding the literature, model reduction problem of LPV-SS representations was investigated in several papers \cite{farhood2003, dehillerin2011, adegas2013, wood1996, widowati}, but except \cite{widowati} they are only applicable to quadratically stable LPV systems. The method of \cite{widowati} is applicable to quadratically stabilizable and detectable LPV-SS representations. In contrast, this paper does not impose any restrictions on the class of LPV-SS representations. 
In \cite{SirajCDC2012} joint reduction of the number of states and the number of scheduling parameters  has been investigated. However, the method of \cite{SirajCDC2012} requires
constructing the Hankel matrix explicitly. Hence, it suffers from the same curse of dimensionality as \cite{toth2012}. In addition, the system theoretic
interpretation of the algorithm is less clear. 
To sum up, the main advantages of the proposed  model reduction algorithm are the following:
\begin{itemize}
\item it is applicable to arbitrary LPV-SS representations,
\item it has a clear system theoretic interpretation,
\item its computational (time and memory) complexity is polynomial in the number of states.
\end{itemize}
The main disadvantage of the presented method is the lack of analytic error bounds. Note, however, that even for classical linear systems, there exists no
analytical error bounds for model reduction algorithms which are based on moment matching.


\textbf{Outline:}
In Section \ref{sect:ALPV_def}, we present the formal definition and main properties of LPV-SS representations. In Section \ref{sect:mod_red_prem}, we recall the concept of sub-Markov parameters for LPV-SS representations and give the precise problem statement. In Section \ref{sect:mod_red}, we present the moment matching algorithm. 
In Section \ref{sect:exam} the algorithm is illustrated on numerical examples and its performance is compared with the one of \cite{toth2012}. 


%

\section{DISCRETE-TIME LPV-SS REPRESENTATIONS}
\label{sect:ALPV_def}

In this section, we present the formal definition of discrete-time LPV-SS representations and recall a number of relevant definitions. We follow the presentation of \cite{petreczky_mercere}.

In the sequel, we will use
\begin{equation} \label{eq:shorthand}
\Sigma=(n_y,n_u,n_x,\{(A_i,B_i,C_i)\}_{i=0}^{n_p}), 
\end{equation}
or simply $\Sigma$ to denote a discrete-time LPV-SS representation of the form \eqref{eq:alpv_int_1}. 
In addition, we use $\mathbb{I}_{s_1}^{s_2}$ to denote the set $\mathbb{I}_{s_1}^{s_2}=\{ s \in \mathbb{N} \mid s_1 \leq s \leq s_2 \}$. An LPV-SS representation $\Sigma$ is driven by the \emph{inputs} $\{ u(k) \}_{k=0}^{\infty}$ and the \emph{scheduling sequence} $\{ p(k) \}_{k=0}^{\infty}$. In the sequel, regarding state trajectories, the initial state $x(0)$ for an LPV-SS representation is taken to be zero unless stated otherwise. This assumption is made to simplify notation. Note that the results of the paper can easily be extended for the case of non-zero initial state.

\begin{Notation}
We will use $H^\mathbb{N}$ to denote the set of all maps of the form $f: \mathbb{N} \rightarrow H$ where $H$ is a (possibly infinite) set. Using this, the sets $\mathcal{U}$, $\mathcal{P}$, $\mathcal{Y}$ and $\mathcal{X}$ are defined as $\mathcal{U}={U}^{\mathbb{N}}$, $\mathcal{P}={P}^{\mathbb{N}}$, $\mathcal{Y}={Y}^{\mathbb{N}}$ and $\mathcal{X}={X}^{\mathbb{N}}$ where $U=\mathbb{R}^{n_u}$, $P=\mathbb{P} \subseteq \mathbb{R}^{n_p}$, $Y=\mathbb{R}^{n_y}$ and $X=\mathbb{R}^{n_x}$. 
\end{Notation} 

 
Consider an initial state $x_0 \in \mathbb{R}^{n_x}$ of the LPV-SS representation $\Sigma$ of the form \eqref{eq:alpv_int_1}. The \emph{input-to-state map} $X_{\Sigma,x_0}: \mathcal{U} \times \mathcal{P} \rightarrow \mathcal{X}$ and \emph{input-output map} $Y_{\Sigma,x_0}: \mathcal{U} \times \mathcal{P} \rightarrow \mathcal{Y}$ of $\Sigma$ corresponding to this initial state $x_0$ are defined as follows: 
for all sequences $\textbf{u}=\{ u(k) \}_{k=0}^{\infty} \in \mathcal{U}$ and $\textbf{p}=\{ p(k) \}_{k=0}^{\infty} \in \mathcal{P}$, let
$X_{\Sigma,x_0}(\textbf{u},\textbf{p})(t)=x(t)$ and $Y_{\Sigma,x_0}(\textbf{u},\textbf{p})(t)=y(t)$, $t  \in \mathbb{N}$,  where
$x(t)$, $y(t)$ satisfy \eqref{eq:alpv_int_1} and $x(0)=x_0$.  
In the sequel, we will use $X_{\Sigma}$ and $Y_{\Sigma}$ to denote $X_{\Sigma,0}$ and $Y_{\Sigma,0}$ respectively.
That is, $X_{\Sigma}$ and $Y_{\Sigma}$ denote the input-to-state and input-output maps which are induced by the zero initial state. In fact, in the sequel we will
be dealing with those input-output maps of LPV-SS representations which correspond to the zero initial state. 
 
The definition above implies that the potential input-output behavior of an LPV-SS representation can be formalized as a map
\begin{equation} \label{eq:f}
	f: \mathcal{U} \times \mathcal{P} \rightarrow \mathcal{Y}.
\end{equation}
The value $f(\mathbf{u},\mathbf{p})(t)$ represents the output of the underlying black-box system at time $t$, if the initial state $x(0)=0$, the input $\textbf{u}= \{ u(k) \}_{k=0}^{\infty}$ and the scheduling sequence $\textbf{p}= \{ p(k) \}_{k=0}^{\infty}$ are fed to the system. Note that this black-box system may or may not admit a realization (description) by an LPV-SS representation, but the input-output behavior of any LPV-SS can be represented by a function of the form \eqref{eq:f}. Next, we define when an LPV-SS representation realizes (describes) $f$.
	The LPV-SS representation $\Sigma$ of the form \eqref{eq:alpv_int_1} is a \emph{realization} of a map $f$ of the form \eqref{eq:f}, if $f$ equals the input-output map of $\Sigma$, i.e., $f=Y_{\Sigma}$.
Two LPV-SS representations $\Sigma_1$ and $\Sigma_2$ are said to be \emph{input-output equivalent} if $Y_{\Sigma_1}=Y_{\Sigma_2}$.
	Let $\Sigma$ be an LPV-SS representation of the form \eqref{eq:alpv_int_1}. We say that $\Sigma$ is \emph{reachable}, if
$\mathbb{R}^{n_x}=\SPAN \{ X_{\Sigma}(\textbf{u},\textbf{p})(t) \mid (\textbf{u},\textbf{p}) \in \mathcal{U} \times \mathcal{P}, t \in \mathbb{N} \}$, i.e.
 $\mathbb{R}^{n_x}$ is the smallest vector space containing all the states which are reachable from $x(0)=0$ by some scheduling sequence and input sequence at some time instance $t$, where $t \in \mathbb{N}$.
%
We say that $\Sigma$ is \emph{observable} if for any two initial states $x_1, x_2 \in \mathbb{R}^{n_x}$, $Y_{\Sigma,x_1}=Y_{\Sigma,x_2}$ implies $x_1=x_2$.
That is, if any two distinct initial states of an observable $\Sigma$ are chosen, then for \emph{some} input and scheduling sequence, the resulting outputs will be different.

	Consider an LPV-SS representation  $\Sigma_1$ of the form \eqref{eq:alpv_int_1} and an LPV-SS representation $\Sigma_2$ of the form
	\begin{equation*}
		\Sigma_2=(n_y,n_u,n_x, \{ (A_i^a,B_i^a,C_i^a) \}_{i=0}^{n_p}).
	\end{equation*}
	A nonsingular matrix $\mathcal{S} \in \mathbb{R}^{n_x \times n_x}$ is said to be an \emph{LPV-SS isomorphism} from $\Sigma_1$ to $\Sigma_2$, if for all $i \in \mathbb{I}_{0}^{n_p}$
	\begin{equation}
		A_i^a\mathcal{S}=\mathcal{S}A_i, \mbox{ } B_i^a=\mathcal{S}B_i, \mbox{ } C_i^a\mathcal{S}=C_i.
	\end{equation}
	In this case $\Sigma_1$ and $\Sigma_2$ are called \emph{isomorphic} LPV-SS representations.
%
	The \emph{order} of $\Sigma$, denoted by $\dim(\Sigma)$ is the dimension of its state-space. That is, if $\Sigma$ is of the form \eqref{eq:alpv_int_1}, then  $\dim(\Sigma)=n_x$.
	Let $f$ be an input-output map of the form \eqref{eq:f}. An LPV-SS realization $\Sigma$ is a \emph{minimal realization of} $f$, if $\Sigma$ is a realization of $f$, and for any LPV-SS representation $\bar{\Sigma}$ which is also a realization of $f$, $\dim(\Sigma) \leq \dim(\bar{\Sigma})$. We say that $\Sigma$ is \emph{minimal}, if $\Sigma$ is a minimal realization of its own input-output map $Y_{\Sigma}$.
From \cite{petreczky_mercere}, it follows that an LPV-SS representation $\Sigma$ is minimal if and only if it is reachable and observable. 
In addition, if two minimal LPV-SS realizations are input-output equivalent, then they are isomorphic.
Note that we defined minimality and input-output equivalence in terms of the input-output map induced by the zero initial state, hence we disregard autonomous dynamics.

\section{MODEL REDUCTION OF LPV-SS REPRESENTATIONS: PRELIMINARIES} \label{sect:mod_red_prem}

In this section, the sub-Markov parameters of a realizable input-output map $f$ and its corresponding LPV-SS representation $\Sigma$ will be defined, and the moment matching problem for LPV-SS realizations will be stated formally. To this end, we recall the concepts of an \emph{infinite impulse response (IIR)} representation of an input-output map \cite{toth2012} and the concept of sub-Markov parameters.

	Consider an LPV-SS representation $\Sigma$ of the form \eqref{eq:alpv_int_1}, and consider its input-output map $f=Y_{\Sigma}$. 
        Recall from \cite{toth2012} that for any input sequence $\textbf{u}=\{u(k)\}_{k=0}^{\infty}$ and scheduling sequence 
        $\textbf{p}=\{p(k)\}_{k=0}^{\infty}$, 
	\begin{equation} \label{eq:IIR}
		f(\textbf{u},\textbf{p})(t)=Y_{\Sigma}(\textbf{u},\textbf{p})(t)=\sum_{m=0}^{t}(h_m \diamond p)(t)u(t-m)
	\end{equation}
	for all $t \in \mathbb{N}$ where 
	\begin{equation} 
       \label{eq:Markov}
	\begin{split}
	& (h_0 \diamond p)(t) = 0, \mbox{ } (h_1 \diamond p)(t)= C(p(t))B(p(t-1)), \\
	& \forall m > 1: (h_m \diamond p)(t) = \\ 
        & C(p(t))A(p(t-1)) \cdots A(p(t-m+1)) B(p(t-m)).
	\end{split}
	\end{equation}
        The representation above is called the IIR of $f=Y_{\Sigma}$. 
From \eqref{eq:Markov} and \eqref{eq:alpv_int_1_add}, it can be seen that the terms $(h_m \diamond p)(t)$, $m \geq 0$ can be written as follows:
\begin{equation}
       \label{eq:Markov2}
	\begin{split}
	& (h_0 \diamond p)(t) = 0, \mbox{ } \\
        & (h_1 \diamond p)(t)= \sum_{q=0}^{n_p} \sum_{q_0=0}^{n_p} C_qB_{q_0}p_q(t)p_{q_0}(t-1) \\
	& (h_m \diamond p)(t)= \\ & \sum_{q=0}^{n_p} \sum_{j_1=0}^{n_p} \cdots \sum_{j_{m-1}=0}^{n_p} \sum_{q_0=0}^{n_p} C_qA_{j_1} \cdots A_{j_{m-1}} B_{q_0} \hat{p}_{qj_1\cdots j_{m-1}q_0}
	\end{split}
\end{equation}
where $p_0(k)=1$ for all $k \in \mathbb{I}_{0}^{t}$ and $\hat{p}_{qj_1 \cdots j_{m-1}q_0}=p_q(t)p_{j_1}(t-1) \cdots p_{j_{m-1}}(t-m+1) p_{q_0}(t-m)$. 

Now we are ready to define the sub-Markov parameters of $\Sigma$. To this end,
we introduce the symbol $\epsilon$ to denote the empty sequence of integers, i.e. $\epsilon$ will stand for a sequence of length zero and we denote
by $\mathcal{S}(\mathbb{I}_{0}^{n_\mathrm{p}})$ the set $\{\epsilon \} \cup \{ j_1\cdots j_m \mid m \ge 1, j_1,\ldots,j_m \in \mathbb{I}_{0}^{n_\mathrm{p}} \}$
of all sequence of integers from  $\mathbb{I}_{0}^{n_\mathrm{p}}$, including the empty sequence. 
If $s \in \mathcal{S}(\mathbb{I}_{0}^{n_\mathrm{p}})$, then $|s|$ denotes the length of the sequence $s$. By convention, if $s=\epsilon$, then
$|s|=0$. 
	The coefficients
        \begin{equation}
        \label{def:Markov:eq1}
        \begin{split}
          & \eta^{\Sigma}_{q,q_0}(\epsilon)=C_qB_{q_0}, \\
          & \eta^{\Sigma}_{q,q_0}(j_1\cdots j_{m})=C_q A_{j_1} \cdots A_{j_{m}} B_{q_0},
       \end{split}
       \end{equation}
 $m \ge 1$; $q,j_1,\dots,j_{m}, q_0 \in \mathbb{I}_{0}^{n_p}$ appearing in \eqref{eq:Markov2} are called the \emph{sub-Markov parameters} of the LPV-SS representation $\Sigma$.  
In the sequel, the sub-Markov parameters $\eta^{\Sigma}_{q,q_0}(s)$, $q,q_0 \in \mathbb{I}_{0}^{n_p}$, $s \in \mathcal{S}(\mathbb{I}_{0}^{n_\mathrm{p}})$, $|s|=m$ will be called \emph{sub-Markov parameters of $\Sigma$ of length $m$}.
The intuition behind this terminology is as follows: the length of a sub-Markov parameter is determined by the number of $A_j$ matrices which appear in \eqref{def:Markov:eq1} as factors. 

Note  the sub-Markov parameters do not depend on the particular choice of an LPV-SS representation, but on the choice of the input-output map (provided that we fix an affine depency of the matrices of the LPV-SS representation on the scheduling variable). 
From \cite{petreczky_mercere} it follows that if $\Sigma_1$, $\Sigma_2$ are
two LPV-SS representations with static affine dependence on the scheduling variable, then their input-output maps are equal, if and only if  their respective sub-Markov parameters are equal, i.e.
\( Y_{\Sigma_1}=Y_{\Sigma_2} \) $\iff$ $ \forall s \in \mathcal{S}(\mathbb{I}_{0}^{n_\mathrm{p}}): \eta_{q,q_0}^{\Sigma_1}(s)=\eta_{q,q_0}^{\Sigma_2}(s)$.
Note also that another way to interpret the sub-Markov parameters is that they correspond to the derivatives of $f$ with respect to the scheduling parameters. 

%
\begin{Example} 
\label{ex:output}
	Let $\Sigma=(n_y,n_u,n_x, \{ (A_i,B_i,C_i) \}_{i=0}^{2})$ be an LPV-SS realization of the map $f=Y_{\Sigma}$. Then the output of $\Sigma$ due to the input $\textbf{u}= \{ u(k) \}_{k=0}^{\infty}$ and scheduling sequence $\textbf{p}= \{ p(k) \}_{k=0}^{\infty}$ at time $t=2$ will be 
	\begin{equation*}
	\begin{split}
	& Y_{\Sigma}(\textbf{u},\textbf{p})(2)= y(2)=\sum\limits_{i=0}^{2}(h_i \diamond p)(2) \cdot u(2-i) \\
	& = 0 + (h_1 \diamond p)(2) \cdot u(2-1) + (h_2 \diamond p)(2) \cdot u(2-2) \\
	& = C(p)B(p(t-1))u(1) + C(p) A(p(t-1)) B(p(t-2))u(0) \\
	& = \sum_{q=0}^{2} \sum_{q_0=0}^{2} C_qB_{q_0}p_q(2)p_{q_0}(1)u(1) \\
	& +  \sum_{q=0}^{2} \sum_{j_1=0}^{2} \sum_{q_0=0}^{2} C_q A_{j_1} B_{q_0}p_q(2) p_{j_1}(1) p_{q_0}(0)u(0).
	\end{split}
	\end{equation*}
\end{Example}

Recall that $p_0(k)=1$ for all $k \in \mathbb{I}_{0}^{t}$. In addition, observe from \eqref{eq:Markov}, that the output $y(t)$, for $t \geq 1$ of an LPV-SS representation corresponding to an input sequence $\textbf{u}= \{ u(k) \}_{k=0}^{\infty}$ and a scheduling sequence  $\textbf{p}= \{ p(k) \}_{k=0}^{\infty}$ is uniquely determined by the sub-Markov parameters of length up to $t-1$ i.e., only the sub-Markov parameters of length up to $t-1$ appear in the output $y(t)$ (see Example \ref{ex:output} for an illustration). Hence, if the sub-Markov parameters of length up to $t-1$ of two LPV-SS representations $\Sigma$ and $\bar{\Sigma}$ coincide, it means that $\Sigma$ and $\bar{\Sigma}$ will have the same input-output behavior up to time $t$ for arbitrary input and scheduling sequences.
This discussion is formalized below.
\begin{Lemma}[I/O equivalence and sub-Markov parameters]
  For any  LPV-SS representations $\Sigma_1,\Sigma_2$, 
 \[ \forall (\mathbf{u}, \mathbf{p}) \in \mathcal{U} \times \mathcal{P}, k \in \mathbb{I}_0^{t}: \quad   Y_{\Sigma_1}(u,p)(k)=Y_{\Sigma_2}(u,p)(k) \]
  if and only if
  \[ 
     \forall s \in \mathcal{S}(\mathbb{I}_{0}^{n_\mathrm{p}}), q,q_0 \in \mathbb{I}_0^{n_{\mathrm{p}}}, |s| \le t-1:  \quad
     \eta_{q,q_0}^{\Sigma_1}(s)=\eta_{q,q_0}^{\Sigma_2}(s)
  \]
\end{Lemma}

This prompts us to introduce the following definition.

\begin{Definition}  \label{def:Npartial}
	Let $\Sigma$ be an LPV-SS representation of the form \eqref{eq:alpv_int_1}.  An LPV-SS representation $\bar{\Sigma}$ of the form \eqref{eq:alpv_int_2} is called a $N$-partial realization of $f=Y_{\Sigma}$, for some $N \in \mathbb{N}$,  if
     \[ \forall s \in \mathcal{S}(\mathbb{I}_{0}^{n_\mathrm{p}}), q,q_0 \in \mathbb{I}_0^{n_{\mathrm{p}}}, |s| \le N: 
     \eta_{q,q_0}^{\Sigma}(s)=\eta_{q,q_0}^{\bar{\Sigma}}(s)
    \]
\end{Definition}
That is, $\bar{\Sigma}$ is an \emph{$N$-partial realization} of $f=Y_{\Sigma}$, if sub-Markov parameters of $Y_{\Sigma}$ and $Y_{\bar{\Sigma}}$ up to length $N$ are equal.  
In other words, 
$\bar{\Sigma}$ is an $N$-partial realization of $Y_{\Sigma}$, if 
	\begin{equation*}
	\begin{split}
        C_qB_{q_0}=\bar{C}_q\bar{B}_{q_0}, \mbox{ } \forall q,q_0 \in \mathbb{I}_{0}^{n_p}, \\
	 C_q A_{j_1} \cdots A_{j_k} B_{q_0}=\bar{C}_q \bar{A}_{j_1} \cdots \bar{A}_{j_k} \bar{B}_{q_0}, \mbox{ }, \forall k \in \mathbb{I}_{1}^{N},  \\
	  \forall q,q_0,j_1, \dots, j_k \in \mathbb{I}_{0}^{n_p}. 
	\end{split}
	\end{equation*}
The problem of model reduction by moment matching for LPV-SS models can now be formulated as follows.
\begin{Problem} \label{prob:momentmatching}
	Let $\Sigma$ be an LPV-SS representation and let $f=Y_{\Sigma}$ be its input-output map. Fix $N \in \mathbb{N}$. Find another LPV-SS realization $\bar{\Sigma}$ such that $\dim (\bar{\Sigma}) < \dim (\Sigma)$ and $\bar{\Sigma}$ is an $N$-partial realization of $f=Y_{\Sigma}$.
\end{Problem}

In order to explain the intuition behind this definition, we combine \cite[Theorem 4]{petreczky} and \cite{petreczky_mercere} to derive the following. 
\begin{Corollary} \label{cor:rel_N_n}
	Assume that $\Sigma$ is a minimal realization of $f=Y_{\Sigma}$ and $N$ is such that $2\dim (\Sigma)-1 \le N$. Then for any LPV-SS representation $\bar{\Sigma}$ which is an $N$-partial realization of $f$, $\bar{\Sigma}$ is also a realization of $f=Y_{\Sigma}$ and $\dim (\Sigma) \le \dim (\bar{\Sigma})$.
\end{Corollary}
\begin{Remark} \label{rem:N_and_n}
	Corollary \ref{cor:rel_N_n} implies that there is a tradeoff between the choice of $N$ and the order of $\Sigma$.
Assume $\Sigma$ is a minimal realization of $f=Y_{\Sigma}$. 
If $N$ is chosen to be too high, namely if it is such that $N \geq 2n_x-1$, then it will not be possible to find an LPV-SS representation which is an $N$-partial realization of $f$ and whose order is lower than $n_x$. In fact, if the model reduction procedure to be presented in the next section is used with any input $N \geq 2n_x-1$, then the resulting LPV-SS representation $\bar{\Sigma}$ will be a complete realization of $f=Y_{\Sigma}$. However, the order of $\bar{\Sigma}$ will be the same as the order of $\Sigma$ (provided that $\Sigma$ is minimal). This relation between $N$ and $n_x$ gives an a priori idea of how well the input-output map of $\bar{\Sigma}$ approximates that of $\Sigma$. More specifically, we can expect the output error $Y_{\Sigma}-Y_{\bar{\Sigma}}$ to be smaller when $N$ is increased, as long as $N < 2n_x-1$. This error will be zero for $N \geq 2n_x-1$, since in this case $\bar{\Sigma}$ will be a complete realization of $Y_{\Sigma}$. 
\end{Remark}

\section{MODEL REDUCTION OF LPV-SS REPRESENTATIONS} \label{sect:mod_red}

In this section, first, the theorems which form the basis of the model reduction by moment matching will be presented. Then the algorithm itself will be stated.
In the sequel, the image (column space) and kernel (null space) of a real matrix $M$ is denoted by $\IM (M)$ and $\ker(M)$ respectively. In addition, $\Rank (M)$ is the dimension of $\IM (M)$. 
We will start with presenting the following definitions for LPV-SS realizations of the form \eqref{eq:alpv_int_1}.
\begin{Definition}[$N$-partial unobservability space] \label{ObservabilityMat}
	The $N$-partial unobservability space $\mathscr{O}_N(\Sigma)$ of $\Sigma$ is defined inductively as follows:
	\begin{equation}
	\begin{aligned}
	& \mathscr{O}_0(\Sigma)= \bigcap_{q \in \mathbb{I}_{0}^{n_p}} \ker(C_q), \\
	& \mathscr{O}_{N}(\Sigma)=\mathscr{O}_{0}(\Sigma) \cap \bigcap_{j \in \mathbb{I}_{0}^{n_p}} \ker (\mathscr{O}_{N-1}(\Sigma)A_j), \mbox{ }  N \geq 1.
	\end{aligned}
	\end{equation}
\end{Definition}
From \cite{MP:BigArticlePartI,petreczky_mercere}, it follows that $\Sigma$ is observable if and only if $\mathscr{O}_{N}(\Sigma)=\{0\}$ for all $N \ge n_x-1$.

\begin{Definition}[$N$-partial reachability space] \label{ReachabilityMat}
	The $N$-partial reachability space $\mathscr{R}_N(\Sigma)$ of $\Sigma$ is defined inductively as follows:
	\begin{equation}
	\begin{aligned}
	& \mathscr{R}_0(\Sigma)= \SPAN \bigcup_{q_0 \in \mathbb{I}_0^{n_p}} \IM(B_{q_0}), \\
	& \mathscr{R}_{N}(\Sigma)=\mathscr{R}_0(\Sigma)+\sum_{j \in \mathbb{I}_{0}^{n_p}} \IM (A_j\mathscr{R}_{N-1}(\Sigma)), \mbox{ } N \geq 1.
	\end{aligned}
	\end{equation}
	where the summation operator must be interpreted as the Minkowski sum.
\end{Definition}
Again, from \cite{MP:BigArticlePartI,petreczky_mercere} it follows that $\Sigma$ is span-reachable if and only if $\dim (\mathscr{R}_N(\Sigma))=n_x$ for all $N \ge n_x-1$.

\begin{Remark} \label{rem:complexity}
        Let $\Sigma$ be a LPV-SS representation of the form \eqref{eq:alpv_int_1}.
        Recall from \cite{toth2012} the definition of the $N$-step extended reachability matrix $R_N$ and the definition of the $N$-step extended observability matrix
        $O_N$ of $\Sigma$. It is easy to see that $\ker (O_N)= \mathscr{O}_N(\Sigma)$ and $\IM(R_N)=\mathscr{R}_N(\Sigma)$. Following \cite{toth2012} define
       Hankel matrix $H_{N,N}$ of an LPV-SS representation $\Sigma$ as $H_{N,N}= O_N R_N$. Note that 
       $H_{N,N}$ is of dimension
	\(
		{\small n_y(n_p+1) \left( \frac{(n_p+1)^{N+1}-1}{n_p} \right) \times n_u(n_p+1) \left( \frac{(n_p+1)^{N+1}-1}{n_p} \right). }
	\)
         i.e. it is exponential in $N$. 
Recall that \cite{toth2012} proposes a Kalman-Ho like realization algorithm based on the factorization of $H_{N,N}$ for some $N$. The problem with this approach is that it involves explicit construction of Hankel matrices. 
Consequently, in the worst-case, memory-usage and time complexity of the algorithm \cite{toth2012} are exponential $N$. 
In \cite{toth2012}, $N$ is chosen so that rank of $H_{N,N}$ equals some integer $n$ and the order of the LPV-SS computed from $H_{N,N}$ will be at most $n$. 
While for many example, $N$ will be small, it can happen that $N$ is large, with $N=n-1$ being the worst-case scenario, see Section \ref{sect:exam} for an example.
In addition, the method in \cite{toth2012} does not solve Problem \ref{prob:momentmatching}, instead it relies on an approximation which is similar to balanced truncation. It yields an LPV-SS representation whose sub-Markov parameters are \emph{close} to the corresponding sub-Markov parameters of the original LPV-SS representation. In Section \ref{sect:exam}, these remarks will be illustrated by numerical examples.
\end{Remark}

\begin{Theorem} \label{theo:mert1}
	Let
	\(
	\Sigma=(n_y,n_u,n_x,\{(A_i,B_i,C_i)\}_{i=0}^{n_p})
	\)
	be an LPV-SS representation, let $V \in \mathbb{R}^{n_x \times r}$ be a full column rank matrix such that
	\[
	\mathscr{R}_{N}(\Sigma) = \IM (V).
	\]
	If $\bar{\Sigma}=(n_y,n_u,r,\{(\bar{A}_i,\bar{B}_i,\bar{C}_i)\}_{i=0}^{n_p})$ is an LPV-SS representation such that for each $i \in \mathbb{I}_{0}^{n_p}$, the matrices $\bar{A}_i,\bar{B}_i,\bar{C}_i$ are defined as
	\[
	\bar{A}_i=V^{-1}A_iV \mbox{, } \bar{B}_i=V^{-1}B_i \mbox{, } \bar{C}_i=C_iV,
	\]
	where $V^{-1}$ is a left inverse of $V$, then $\bar{\Sigma}$ is an $N$-partial realization of the input-output map $f=Y_{\Sigma}$ of $\Sigma$.
\end{Theorem}

This theorem follows from \cite{bastugACC2014}, \cite{bastug_IEEE_TAC_2015_arxiv} using \cite{petreczky_mercere}. For the sake of completeness, we present the proof below.

\begin{proof}
Let $N=0$. Since the conditions of Theorem~\ref{theo:mert1} imply $\IM (B_{q_0}) \subseteq \IM (V)$, $q_0 \in \mathbb{I}_{0}^{n_p}$ and $V^{-1}$ is a left inverse of $V$, it is a routine exercise to see that $VV^{-1}B_{q_0}=B_{q_0}$. If $N \geq 1$, then $\IM (A_{j_i} \cdots A_{j_1}B_{q_0})$ is also a subset of $\mathscr{R}_N(\Sigma)=\IM (V)$, $i=1,\ldots,N$. Hence, by induction we can show that 
$VV^{-1}A_{j_i} \cdots A_{j_1}B_{q_0}=A_{j_i} \cdots A_{j_1}B_{q_0}$, $i=1,\ldots,N$, which ultimately yields
\begin{equation} \label{eq2:theo_mert1}
V\bar{A}_{j_{N}}\cdots \bar{A}_{j_1}\bar{B}_{q_0}=A_{j_{N}} \cdots A_{j_1}B_{q_0}.
\end{equation}
Using \eqref{eq2:theo_mert1}, and $\bar{C}_q=C_qV$, $q \in \mathbb{I}_{0}^{n_p}$, we conclude that for all $i \le N$;
$q,q_0,j_1, \dots, j_i \in \mathbb{I}_{0}^{n_p}$, 
\[ 
\bar{C}_q \bar{A}_{j_i} \cdots \bar{A}_{j_1} \bar{B}_{q_0}=C_q A_{j_i} \cdots A_{j_1} B_{q_0}
\]
from which the statement of the theorem follows.
\end{proof}

Note that the number $r$ is the number of columns in the full column rank matrix $V$, hence $r \leq n_x$. This fact leads $\bar{\Sigma}$ to be of reduced order if $N$ is sufficiently small, see Corollary \ref{cor:rel_N_n}. Using a dual argument, we can prove the following.

\begin{Theorem}
\label{theo:mert2}
	Let $\Sigma=(n_y,n_u,n_x,\{(A_i,B_i,C_i)\}_{i=0}^{n_p})$ be an LPV-SS representation, and let
     $W \in \mathbb{R}^{r \times n_x}$ be a full row rank matrix such that
	\[
	\mathscr{O}_{N}(\Sigma) = \ker (W).
	\]
	Let $W^{-1}$ be any right inverse of $W$ and let
	\[
	\bar{\Sigma}=(n_y,n_u,r,\{(\bar{A}_i,\bar{B}_i,\bar{C}_i)\}_{i=0}^{n_p})
	\]
	be an LPV-SS representation such that for each $i \in \mathbb{I}_{0}^{n_p}$, the matrices $\bar{A}_i,\bar{B}_i,\bar{C}_i$ are defined as
	\[
	\bar{A}_i=WA_iW^{-1} \mbox{, } \bar{B}_i=WB_i \mbox{, } \bar{C}_i=C_iW^{-1}.
	\]
	Then $\bar{\Sigma}$ is an $N$-partial realization of the input-output map $f=Y_{\Sigma}$ of $\Sigma$.
\end{Theorem}
The proof is similar to that of Theorem \ref{theo:mert1}.

Finally, by combining the proofs of Theorem \ref{theo:mert1} and Theorem \ref{theo:mert2}, we can show the following.

\begin{Theorem} 
	\label{theo:mert3}
	Let $\Sigma=(n_y,n_u,n_x,\{(A_i,B_i,C_i)\}_{i=0}^{n_p})$ be an LPV-SS representation, and let $V \in \mathbb{R}^{n_x \times r}$ and $W \in \mathbb{R}^{r \times n_x}$ be respectively full column rank and full row rank matrices such that
	\[
	\mathscr{R}_{N}(\Sigma) = \IM (V) \mbox{, } \mathscr{O}_{N}(\Sigma) = \ker (W) \mbox{ and } \Rank(WV)=r.
	\]
	If $\bar{\Sigma}=(n_y,n_u,r,\{(\bar{A}_i,\bar{B}_i,\bar{C}_i)\}_{i=1}^{n_p})$ is an LPV-SS representation such that for each $i \in \mathbb{I}_{0}^{n_p}$,  $\bar{A}_i,\bar{B}_i,\bar{C}_i$ are defined as
	\[
	\bar{A}_i=WA_iV(WV)^{-1} \mbox{, } \bar{B}_i=WB_i \mbox{, } \bar{C}_i=C_iV(WV)^{-1}
	\]
	then $\bar{\Sigma}$ is a $2N$-partial realization of the input-output map $f=Y_{\Sigma}$ of $\Sigma$.
\end{Theorem}

Note that having a $2N$-partial realization as an approximation realization would be more desirable than having an $N$-partial realization, since number of matched sub-Markov parameters would increase. However, it is only possible to get a $2N$-partial realization for the original model $\Sigma$ when the additional condition $\Rank(V)=\Rank(W)=\Rank(WV)=r$ is satisfied. 

Now, we will present an efficient algorithm of model reduction by moment matching, which computes either an $N$ or $2N$-partial realization $\bar{\Sigma}$ for an $f$ which is realized by an LPV-SS representation $\Sigma$. First, we present algorithms for computing the subspaces $\mathscr{R}_N(\Sigma)$ and $\mathscr{O}_N(\Sigma)$. To this end, we will use the following notation: if $M$ is any real matrix, then denote by $\mathbf{orth}(M)$ the matrix $U$ such that $U$ is full column rank, $\IM (U)=\IM (M)$ and $U^{\mathrm{T}}U=I$. Note that $U$ can easily be computed from $M$ numerically, see for example the Matlab command \texttt{orth}.

The algorithm for computing $V  \in \mathbb{R}^{n_x \times r}$ such that $\IM(V)=\mathscr{R}_N(\Sigma)$ is presented in Algorithm \ref{alg1} below.

\begin{algorithm}[h]
	\caption{
		Calculate  a matrix representation of $\mathscr{R}_N(\Sigma)$,
		\newline
		\textbf{Inputs}: $(\{A_i,B_i\}_{i \in \mathbb{I}_{0}^{n_p}})$ and $N$
		\newline
		\textbf{Outputs:} $V  \in \mathbb{R}^{n_x \times r}$ such that $\Rank (V)=r$,
		$\IM (V) = \mathscr{R}_N(\Sigma)$.
	}
	\label{alg1}
	\begin{algorithmic}
		\STATE $V:=U_0$, $U_0:=\mathbf{orth}\begin{bmatrix} B_0 & \cdots & B_{n_p} \end{bmatrix}$.
		\FOR{$k=1\ldots N$} 
		\STATE
		$V:=\mathbf{orth}(\begin{bmatrix} V & A_0V & A_1V & \cdots & A_{n_p}V \end{bmatrix})$
		\ENDFOR
		\RETURN $V$.
	\end{algorithmic}
\end{algorithm}
By duality, we can use Algorithm \ref{alg1} to compute a $W \in \mathbb{R}^{r \times n_x}$ such that $\ker(W)=\mathscr{O}_N(\Sigma)$, see Algorithm \ref{alg2}.

\begin{algorithm}
	\caption{
		Calculate a matrix representation of $\mathscr{O}_N(\Sigma)$
		\newline
		\textbf{Inputs}: $\{A_i,C_i\}_{i \in \mathbb{I}_{0}^{n_p}}$ and $N$
		\newline
		\textbf{Output:} $W \in \mathbb{R}^{r \times n_x}$, such that
		$\Rank (W) = r$, and $\ker (W)=\mathscr{O}_N(\Sigma)$.
	}
	\label{alg2}
	\begin{algorithmic}
		\STATE Apply Algorithm \ref{alg1} with inputs $(\{A_i^{\mathrm{T}},C_i^{\mathrm{T}}\}_{i \in \mathbb{I}_{0}^{n_p}})$ to obtain
		a matrix $V$.
		\RETURN $W=V^{\mathrm{T}}$.
	\end{algorithmic}
\end{algorithm} 
Notice that the computational complexity of Algorithm \ref{alg1} and Algorithm \ref{alg2} is polynomial in $N$ and $n_x$, even though the spaces of $\mathscr{R}_N(\Sigma)$ (resp. $ \mathscr{O}_N(\Sigma)$) are generated by images (resp. kernels) of exponentially many matrices.
Using Algorithms \ref{alg1} and \ref{alg2}, we can formulate a model reduction algorithm, see Algorithm \ref{alg3}.
\begin{algorithm}
	\caption{Moment matching for LPV-SS representations
		\newpage 
		\textbf{Inputs:} $\Sigma=(n_y,n_u,n_x,\{(A_i,B_i,C_i)\}_{i=0}^{n_p})$, $\texttt{Mode} \in \{ \texttt{R},\texttt{O},\texttt{T} \}$ and $N \in \mathbb{N}$.
		\newpage  
		\textbf{Output: } $\bar{\Sigma}=(n_y,n_u,r,\{(\bar{A}_i,\bar{B}_i,\bar{C}_i)\}_{i=0}^{n_p})$.
	}
	\label{alg3}
	\begin{algorithmic}
		\STATE Using Algorithm \ref{alg1}-\ref{alg2} compute matrices $V$ and $W$ such that
		$V$ is full column rank, $W$ is full row rank and $\IM (V)= \mathscr{R}_N(\Sigma)$,
		$\ker (W) = \mathscr{O}_N(\Sigma)$.
		\IF{$\Rank (V)=\Rank (W)=\Rank (WV)$ and $\texttt{Mode}=\texttt{T}$} 
		\STATE
		Let $r=\Rank (V)$ and
		\begin{align*}
		& \bar{A}_i=WA_iV(WV)^{-1} \mbox{, } \bar{C}_i=C_iV(WV)^{-1} \mbox{, } \\
		& \bar{B}_i=WB_i. 
		\end{align*}
		\ENDIF
		\IF{$\texttt{Mode}=\texttt{R}$}
		\STATE
		Let $r=\Rank (V)$, $V^{-1}$ be a left inverse of $V$ and set
		\[
		\bar{A}_i=V^{-1}A_iV \mbox{, } \bar{C}_i=C_iV \mbox{, } \bar{B}_i=V^{-1}B_i.
		\]
		\ENDIF
		\IF{$\texttt{Mode}=\texttt{O}$}
		\STATE
		Let $r=\Rank (W)$ and let $W^{-1}$ be a right inverse of $W$. Set
		\[
		\bar{A}_i=WA_iW^{-1} \mbox{, } \bar{C}_i=C_iW^{-1} \mbox{, } \bar{B}_i=WB_i.
		\]
		\ENDIF
		\RETURN $\bar{\Sigma}=(n_y,n_u,r,\{(\bar{A}_i,\bar{B}_i,\bar{C}_i)\}_{i=0}^{n_p})$.
	\end{algorithmic}
\end{algorithm}

Theorems \ref{theo:mert1} -- \ref{theo:mert3} imply the correctness of Algorithm \ref{alg3}.

\begin{Corollary}
	Using the notation of Algorithm \ref{alg3}, the following holds: If $\Rank (V)=\Rank (W) = \Rank (WV)$ and $\texttt{Mode}=\texttt{T}$, then Algorithm \ref{alg3} returns a $2N$-partial realization of $f=Y_{\Sigma}$ (if $\texttt{Mode}=\texttt{T}$ and the rank condition does not hold, the algorithm returns nothing). Otherwise, Algorithm \ref{alg3} returns an $N$-partial realization of $f=Y_{\Sigma}$.
\end{Corollary}

Note that even if the condition $\Rank (V)=\Rank (W)=\Rank (WV)$ does not hold, Algorithm \ref{alg3} can always be used for getting an $N$-partial realization, by choosing $\texttt{Mode}=\texttt{O}$ or  $\texttt{Mode}=\texttt{R}$.


\begin{Remark}[Minimization of LPV-SS representations]
	From \cite{petreczky_mercere}, it follows that if $N \geq n_x-1$ then
	\begin{equation*}
	\begin{aligned}
	& \mathscr{R}_N(\Sigma)= \sum_{i=0}^{\infty} \mathscr{R}_i(\Sigma)= \\ & \SPAN \{ X_{\Sigma}(\textbf{u},\textbf{p})(t) \mid (\textbf{u},\textbf{p}) \in \mathcal{U} \times \mathcal{P}, t \geq 0 \}, \\
	& \mathscr{O}_N(\Sigma)= \bigcap_{i=0}^{\infty} \mathscr{O}_i(\Sigma) = \\ &  \{ x \in \mathbb{R}^{n_x} \mid Y_{\Sigma,x}(\textbf{u},\textbf{p})(t)=0, \forall (\textbf{u},\textbf{p}) \in \mathcal{U} \times \mathcal{P}, \forall t \geq 0 \}.
	\end{aligned}
	\end{equation*}
	In other words, an LPV-SS representation $\Sigma$  of the form \eqref{eq:alpv_int_1} is reachable if and only if the dimension of its $N$-partial reachability space $\mathscr{R}_{N}(\Sigma)$ is $n_x$ for all $N \geq n_x-1$, and $\Sigma$ is observable if and only if the dimension of its $N$-partial unobservability space $\mathscr{O}_{N}(\Sigma)$ is $0$ for all $N \geq n_x-1$. In addition from \cite{petreczky_mercere}, it follows that $\Sigma$ is a minimal realization of its own input-output map $Y_{\Sigma}$ if and only if $\Sigma$ is reachable and observable. Hence, using this fact and \cite{MP:BigArticlePartI}, \cite{toth2012}, it can be shown that Algorithm \ref{alg3} can be used as an order minimization algorithm. That is, Algorithm \ref{alg3} can be used consecutively with the inputs $N \geq n_x-1$, $\texttt{Mode}=\texttt{R}$ (in this case, the resulting $\bar{\Sigma}$ will be reachable and it will be a realization of $f=Y_{\Sigma}$) and $N \geq n_x-1$, $\texttt{Mode}=\texttt{O}$ (in this case, the resulting $\bar{\Sigma}$ will be observable and it will be a  realization of $f=Y_{\Sigma}$) for reachability and observability reduction for $\Sigma$, respectively. In turn, the resulting representation $\bar{\Sigma}$ will be a minimal realization of $f=Y_{\Sigma}$.	
\end{Remark}

\begin{Remark}[Order $r$ of the reduced representation]
	A disadvantage of the model reduction algorithm proposed by this paper is that the order of the reduced model produced by the method is unknown a priori. Namely, only the number $N$ is chosen by the user as an input to the procedure, and the order of the reduced LPV-SS representation for this $N$ is unknown beforehand. However, this issue can easily be solved by slightly modifying the method according to the concept of nice selections \cite{bastug_IEEE_TAC_2015_arxiv}. 
We omit the details of this approach due to lack of space. 
\end{Remark}

\section{NUMERICAL EXAMPLES}
\label{sect:exam}

In this section, initially, the method stated in the present paper is applied to Example $4$ in \cite{toth2012} and the result is compared with the one given in \cite{toth2012}. For this, both procedures are implemented in \textsc{Matlab}. The codes and the data used for both examples in this section are available from https://kom.aau.dk/\texttildelow mertb/.

First, the algorithm is applied to get a $3$rd order approximation to the LPV-SS realization of order $4$ in Example $4$, \cite{toth2012}. The original LPV-SS representation used in this case is of the form $\Sigma=(n_y,n_u,n_x,\{(A_i,B_i,C_i)\}_{i=0}^{n_p})$ with $n_y=n_u=1$, $n_x=4$ and $n_p=3$. When $N$ is chosen to be $1$ and $\texttt{Mode}=\texttt{Reach}$, the resulting reduced order model $\bar{\Sigma}$ is a $1$-partial realization of $Y_{\Sigma}$ of order $3$. The scheduling signal used for simulation is of the form $p(t)= \begin{bmatrix} \hat{p} & \sqrt{-\hat{p}} & \sin(\hat{p}) \end{bmatrix}^\mathrm{T}$ where the parameter $\hat{p}$ takes its values randomly at each time instant, in the interval $[-2 \pi, 0]$. In addition, a white input $u(t) \sim \mathcal{N}(0,1)$ is used. The upper limit of the simulation time interval is chosen to be $N+50=51$. Since $N=1$, the sub-Markov parameters of length at most $1$ are matched with the original LPV-SS model $\Sigma$. The precise number of matched sub-Markov parameters is thus:
\begin{equation} \label{eq:Markov_parameter_number}
	(n_p+1) \left( \frac{(n_p+1)^{N+1}-1}{n_p} \right) (n_p+1)= 80 
\end{equation}
The original model $\Sigma$ and the the reduced order model $\bar{\Sigma}$ are simulated for $500$ different scheduling and input signal sequences of the type explained above, and their outputs $y(t)$ and $\bar{y}(t)$ are compared for $t=0,1,\ldots,K$, where $K$ is the number of steps of the simulation. For each simulation, the responses of $\Sigma$ and $\bar{\Sigma}$ are compared with the best fit rate (BFR) (see \cite{ljung}, \cite{toth2012}) which is defined as
\[
\mbox{BFR}=100 \% \max \left( 1-\frac{\sqrt{\sum_{t=0}^{K}\norm{y(t)-\bar{y}(t)}^2_2}}{\sqrt{\sum_{t=0}^{K} \norm{y(t)-y_m}^2_2}},0 \right)
\]
where $y_m$ is the mean of $\{y(t)\}_{t=0}^{K}$. 

\begin{figure}
	\centering
	\includegraphics[width=0.4\textwidth]{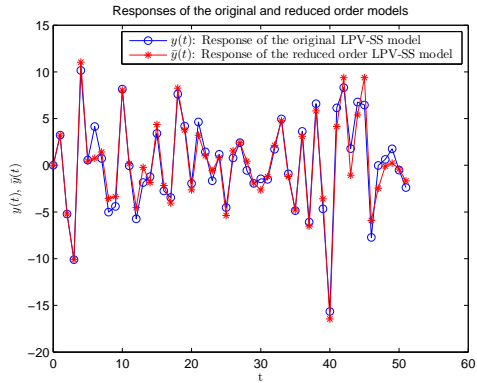}
	\caption{The responses of the original LPV-SS model $\Sigma$ of order $4$ and the reduced order LPV-SS model $\bar{\Sigma}$ of order $3$ acquired by Algorithm~\ref{alg3}. The BFR for this simulation is $=76.5773\%$.}
	\label{fig:example1_toth}
\end{figure}

For this example, the algorithms stated in this paper and in \cite{toth2012} are implemented for comparison. The mean of the BFRs, which is computed over $500$ simulations, can be seen on Table \ref{table1}. In addition the best and worst BFRs over $500$ simulations and the run-times for one single reduction algorithm are also shown in Table \ref{table1}. The outputs $y(t)$ and $\bar{y}(t)$ of the simulation which give the closest value to the mean of the BFRs are shown in Fig.~\ref{fig:example1_toth}.  We used Algorithm \ref{alg3} to perform model reduction using moment matching. 
\begin{table}[!t]
\caption{Comparison of Alg. \ref{alg3} and the Alg. in \cite{toth2012}}
\label{table1}
\centering
\begin{tabular}{|c|c|c|c|c|} 
	\hline
	\textbf{The Proc.} & \textbf{Mean BFR} & \textbf{Best BFR} & \textbf{Worst BFR} & \textbf{Run Time} \\ \hline
	Alg. \ref{alg3} & $76.5710\%$ & $86.5821\%$ & $64.9409\%$ & $0.0430$ s \\ \hline
	Alg. in \cite{toth2012} & $75.4364\%$ & $85.4157 \%$ & $58.5798 \%$ & $0.0711$ s \\
	\hline
\end{tabular}
\end{table}
From Table \ref{table1}, it can be seen that both algorithms result in almost the same fit rates, whereas the algorithm stated in the present paper provides a $50$\%   in  terms of computational complexity. 


Next, a numerical example is presented to further illustrate the difference between the algorithms of the present paper, and the algorithm in \cite{toth2012}. 
The algorithm in the present paper is applied to get a reduced order approximation to a minimal LPV-SS model whose linear subsystems are stable. The original LPV-SS model used in this case is of the form $\Sigma=(n_y,n_u,n_x,\{(A_i,B_i,C_i)\}_{i=0}^{n_p})$ with $n_y=n_u=1$, $n_x=7$ and $n_p=5$. The $\{(A_i,B_i,C_i)\}_{i=0}^{n_p}$ parameters of $\Sigma$ are as follows:
\begin{equation*}\begin{split}
A_0 & = \left[ \begin{array}{ccc} -0.5 & 0.5471 & \textbf{0}_{1 \times 5} \\ \textbf{0}_{6 \times 1} & \textbf{0}_{6 \times 1} & \textbf{0}_{6 \times 5}	\end{array} \right] \\
A_1 & = \left[ \begin{array}{cccc} \textbf{0}_{1 \times 1} & \textbf{0}_{1 \times 1} & \textbf{0}_{1 \times 1} & \textbf{0}_{1 \times 4} \\ 
                                   \textbf{0}_{1 \times 1} & 0.3 & 0.2285 & \textbf{0}_{1 \times 4} \\
                                   \textbf{0}_{5 \times 1} & \textbf{0}_{5 \times 1} & \textbf{0}_{5 \times 1} & \textbf{0}_{5 \times 4}	\end{array} \right] \\
A_2 & = \left[ \begin{array}{cccc} \textbf{0}_{2 \times 1} & \textbf{0}_{2 \times 1} & \textbf{0}_{2 \times 1} & \textbf{0}_{2 \times 3} \\ 
								   \textbf{0}_{1 \times 2} & -0.4 & 0.4741 & \textbf{0}_{1 \times 3} \\
								   \textbf{0}_{4 \times 2} & \textbf{0}_{4 \times 1} & \textbf{0}_{4 \times 1} & \textbf{0}_{4 \times 3}	\end{array} \right] \\
\end{split}
\end{equation*}
\begin{equation*}\begin{split}
A_3 & = \left[ \begin{array}{cccc} \textbf{0}_{3 \times 3} & \textbf{0}_{3 \times 1} & \textbf{0}_{3 \times 1} & \textbf{0}_{3 \times 2} \\ 
								   \textbf{0}_{1 \times 3} & -0.7 & 0.9362 & \textbf{0}_{1 \times 2} \\
								   \textbf{0}_{3 \times 3} & \textbf{0}_{3 \times 1} & \textbf{0}_{3 \times 1} & \textbf{0}_{3 \times 2}	\end{array} \right] \\
A_4 & = \left[ \begin{array}{cccc} \textbf{0}_{4 \times 4} & \textbf{0}_{4 \times 1} & \textbf{0}_{4 \times 1} & \textbf{0}_{4 \times 1} \\ 
								   \textbf{0}_{1 \times 4} & 0.5 & 0.4367 & \textbf{0}_{1 \times 1} \\
								   \textbf{0}_{2 \times 4} & \textbf{0}_{2 \times 1} & \textbf{0}_{2 \times 1} & \textbf{0}_{2 \times 1}	\end{array} \right] \\
A_5 & = \left[ \begin{array}{ccc}  \textbf{0}_{5 \times 5} & \textbf{0}_{5 \times 1} & \textbf{0}_{5 \times 1} \\ 
								   \textbf{0}_{1 \times 5} & 0.1 & 0.0573 \\
								   \textbf{0}_{1 \times 5} & \textbf{0}_{1 \times 1} & \textbf{0}_{1 \times 1} \end{array} \right]
\end{split}
\end{equation*}
\begin{equation*}
\begin{split}
B_0 & = \begin{bmatrix}  0 & 0 & 0 & 0 & 0 & 0 & 1 \end{bmatrix}^{\mathrm{T}} \\
B_1 & = \begin{bmatrix}  0 & 0 & 0 & 0 & 0 & 1 & 0 \end{bmatrix}^{\mathrm{T}} \\
B_2 & = \begin{bmatrix}  0 & 0 & 0 & 0 & 1 & 0 & 0 \end{bmatrix}^{\mathrm{T}} \\
B_3 & = \begin{bmatrix}  1 & 0 & 0 & 0 & 0 & 0 & 0 \end{bmatrix}^{\mathrm{T}} \\
B_4 & = \begin{bmatrix}  0 & 1 & 0 & 0 & 0 & 0 & 0 \end{bmatrix}^{\mathrm{T}} \\
B_5 & = \begin{bmatrix}  0 & 0 & 1 & 0 & 0 & 0 & 0 \end{bmatrix}^{\mathrm{T}} \\
C_i & = \begin{bmatrix}  1 & 0 & 0 & 0 & 0 & 0 & 0 \end{bmatrix}, \forall i \in \mathbb{I}_{0}^{n_p}.
\end{split}
\end{equation*}
where $\textbf{0}_{a \times b}$, $a,b \in \mathbb{N} \backslash \{0\}$ denotes the zero matrix of dimension $a \times b$.

The resulting reduced order model $\bar{\Sigma}$ is a $2$-partial realization (hence $N=2$) of $Y_{\Sigma}$ of order $3$. A random scheduling signal and $u(t) \sim \mathcal{N}(0,1)$ is used for simulation. The upper limit of the simulation time interval is chosen to be $N+50=52$. Since $N=2$, the sub-Markov parameters of length at most $2$ are matched with the original LPV-SS model $\Sigma$. Note that the precise number of matched sub-Markov parameters can be found by using \eqref{eq:Markov_parameter_number} with $n_p=5$, $N=2$, which is in this case $1548$.

\begin{figure}
	\centering
	\includegraphics[width=0.4\textwidth]{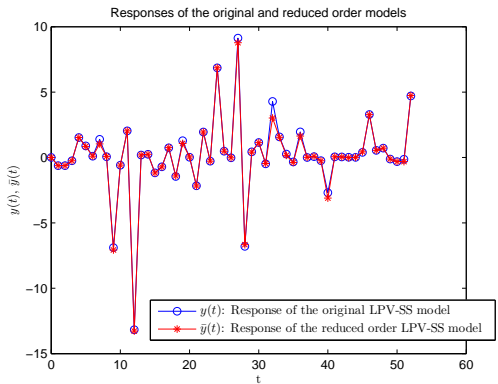}
	\caption{The responses of the original LPV-SS model $\Sigma$ of order $7$ and the reduced order LPV-SS model $\bar{\Sigma}$ of order $3$ acquired by Algorithm~\ref{alg3}. The BFR for this simulation is $=93.4993\%$.}
	\label{fig:example2}
\end{figure}

The output $y(t)$ of the original model $\Sigma$ and the output $\bar{y}(t)$ of the reduced order model $\bar{\Sigma}$ are simulated again for $500$ random scheduling and white Gaussian input signal sequences. For this example, the mean of the BFRs over $500$ simulations is  $93.4888\%$; whereas, the best BFR is $99.3192\%$ and the worst is $47.9013\%$. The elapsed time for all of the simulations is $5.347983$ seconds. The outputs $y(t)$ and $\bar{y}(t)$ of the simulation which gives the closest value to the mean of the BFRs are shown in Fig.~\ref{fig:example2}. It can be seen that the responses of both models are exactly matched until (and including) the time instant $t=N=2$. In addition, Fig.~\ref{fig:example2} together with its BFR$=93.4993\%$ show that the reduced order model $\bar{\Sigma}$ captures the behavior of the original model $\Sigma$ accurately, for the rest of the total time horizon, i.e., for $t > 2$. Note that when the method given in \cite{toth2012} is applied to this example, the method breaks down by running out of memory while trying to compute the smallest Hankel matrix of rank $\dim(\Sigma)=n_x=7$. 


\begin{figure}
	\centering
	\includegraphics[width=0.4\textwidth]{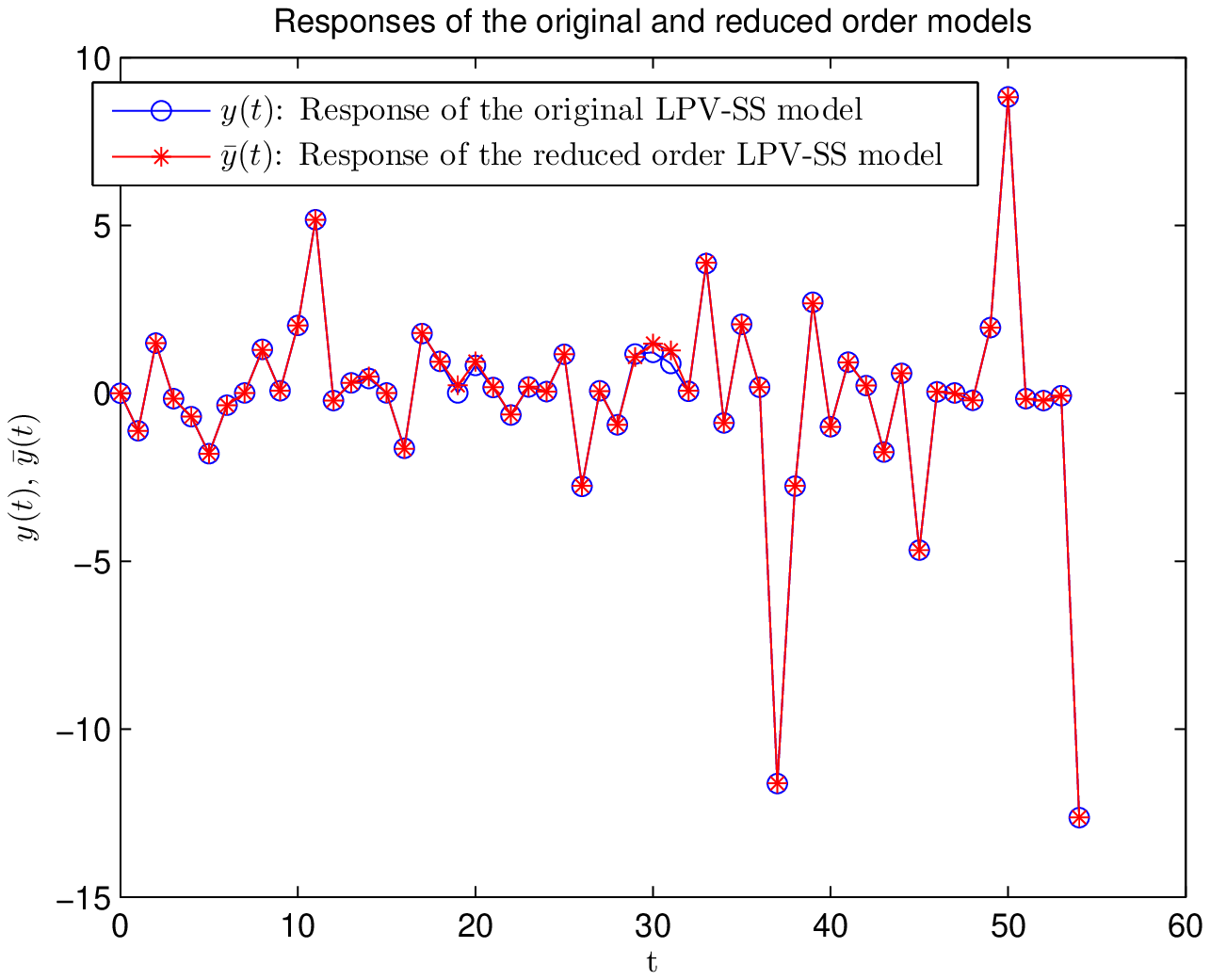}
	\caption{The responses of the original LPV-SS model $\Sigma$ of order $7$ and the reduced order LPV-SS model $\bar{\Sigma}$ of order $5$ acquired by Algorithm~\ref{alg3}. The BFR for this simulation is $=97.4002\%$.}
	\label{fig:example3}
\end{figure}

For the same example, the procedure in the present paper is applied again with $N=4$. Then the resulting reduced order model is of order $5$ and the BFRs over $500$ simulations are as follows: Mean BFR$=97.4010$, Best BFR$=99.8494$, Worst BFR$=75.9829$. The elapsed time for all simulations is $5.621238$ seconds. The outputs $y(t)$ and $\bar{y}(t)$ of the original and the reduced order models which give the closest value to the mean of the BFRs are shown in Fig.~\ref{fig:example3}. Finally, the procedure is applied to get a full realization. For this example, for $N \geq 6$, the reduced LPV-SS representation has the same order with and it is isomorphic to the original LPV-SS representation considered. Hence, it is a full realization of $f=Y_{\Sigma}$. The elapsed time for computing one such full realization for this example is $0.029692$ seconds Note that this is the run-time for only one reduction procedure. No simulations were done to compare the outputs in this case, because they would be exactly the same for all input and scheduling sequences.

\section{CONCLUSIONS}

A model reduction method is presented for discrete time LPV-SS representations with affine static dependence on the scheduling variable. The method makes it possible to find a reduced order approximation to the original LPV-SS model, which has the same input-output behavior for scheduling and input sequences of a pre-defined, limited length. The presented method can also be used for reachability and observability reduction (i.e., minimization) for LPV-SS models.

\bibliographystyle{plain}
\bibliography{./cdc2015}

\addtolength{\textheight}{-12cm}   

\end{document}